\documentclass[12pt,a4paper]{article}
\usepackage{amsmath}
\usepackage[dvips]{graphicx}
\input epsf
\usepackage{times}

\begin{document}
\begin{titlepage}
\title{Sum rules for spin asymmetries}
\author{S.M. Troshin, N.E. Tyurin \\[1ex]
\small  \it Institute for High Energy Physics,\\
\small  \it Protvino, Moscow Region, 142280, Russia} \normalsize
\date{}
\maketitle
\begin{abstract}
Starting from rotational invariance we derive  sum rules for the
single--spin asymmetries in inclusive production and binary
processes.  We also get sum rules for spin correlation parameters
in elastic $pp$--scattering.
\end{abstract}
\end{titlepage}
An important role of  spin effects for analysis of hadron
interaction dynamics is widely recognized nowadays. The
space--time structure of the strong interactions provides a number
of constraints for the spin observables (cf. \cite{bks}) and, as
it will be shown further, allows us to get a useful sum rule
 for the single--spin asymmetries and spin correlation parameters.

Let us consider first single--spin asymmetry in hadron production
\[
h_1+h_2\to h_3+X,
\]
where  the beam or target hadron $h_{1,2}$ is transversely
polarized. Let $\xi$ stands for the set of variables related to
the hadron $h_3$. The definition of asymmetry $A_N$ is well known
\[ A_N(s,\xi)=
\left[\frac{d\sigma^\uparrow}{d\xi}(s,\xi)-\frac{d\sigma^\downarrow}{d\xi}(s,\xi)\right]/
\left[\frac{d\sigma^\uparrow}{d\xi}(s,\xi)+\frac{d\sigma^\downarrow}{d\xi}(s,\xi)\right].
\]
  The equality of the integrated
inclusive cross-sections follows from rotational invariance in a
straightforward way, i.e.
\[
\int
\frac{d\sigma^\uparrow}{d\xi}(s,\xi)d\xi=\int  \frac{d\sigma^\downarrow}{d\xi}(s,\xi)d\xi.
\]
Then from the definition of $A_N$ we have the following sum rule
\begin{equation}\label{srincl}
\int A_N(s,\xi)\frac{d\sigma}{d\xi}(s,\xi)d\xi=0,
\end{equation}
where ${d\sigma}/{d\xi} $ is the unpolarized cross-section. Eq.
(\ref{srincl}) should be taken into account at the construction of
models intended to explain the significant single--spin
asymmetries observed in the inclusive processes. Similar sum rule
(with replacement $A_N$ $\to$ $P$) takes place when the
polarization of the final hadron $h_3$ can be measured
($\Lambda$--hyperon for example).

The above arguments can be  applied to analyzing power in elastic
and binary processes, e.g.  from the equality
\[
\sigma_{el}^\uparrow(s)=\sigma_{el}^\downarrow(s)
\]
we should have
\begin{equation}\label{srelas}
\int_{-s+4m^2}^0 A(s,t)\frac{d\sigma}{dt}(s,t)dt=0,
\end{equation}
where $A(s,t)$ is the analyzing power in elastic scattering and
${d\sigma}/{dt}$ is the unpolarized cross--section for the elastic
scattering of the particles with equal masses. Sum rule for
the inelastic binary processes has the similar form  with minor
kinematical changes of the integration limits. {\it From Eq.
(\ref{srelas}) we arrive to conclusion that $A(s,t)$ should have
sign--changing $t$-dependence since ${d\sigma}/{dt}$ is positive}.
This conclusion on the $t$--dependence is useful for the planning
of the future experiments on the analyzing power measurements in
elastic scattering at higher values of $t$ \cite{adk}. It should
be noted that change of sign of $A$ in elastic $pp$--scattering
was revealed for the first time in the measurements at 40 $GeV/c$
\cite{kaz} and considered as a new experimental regularity in the
analyzing power $t$--dependence that time.

 Oscillating pattern of analyzing power $t$-dependence with
amplitude of oscillations increasing with $t$ observed in various
experiments in elastic and binary processes \cite{bin}, is in
conformity with the sum rule Eq. (\ref{srelas}). Such oscillating
dependence has obtained  model explanation in \cite{osa}. It
should be noted that the Eq. (\ref{srincl}) does not imply similar
$p_{\perp}$--dependence for the single--spin asymmetries in the
inclusive processes and the corresponding experimental data have
not revealed oscillations (cf. e.g. \cite{bks}).

Using rotational invariance combined with particle identity we can
obtain similar sum rules for the spin correlation parameters in
elastic and inclusive $pp$--scattering. Spin correlation
parameters are the spin observables which describe dependence of
the interaction on the relative orientations of the spins of the
two particles (cf. \cite{bks}). We will consider scattering when
both protons in {\it the  initial state} are polarized.
Definition of spin correlation parameter $A_{nn}$ is the following
\begin{equation}\label{ann}
A_{nn}=\frac{{\frac{d\sigma_{^{\uparrow\uparrow }} }{dt}} +
{\frac{d\sigma_{^{\downarrow\downarrow }} }{dt}} -
{\frac{d\sigma_{^{\uparrow\downarrow }}}{dt}} -
{\frac{d\sigma_{^{\downarrow\uparrow }}}{dt}}}
{{\frac{d\sigma_{^{\uparrow\uparrow }}}{dt}}
 +{\frac{d\sigma_{^{\downarrow\downarrow }} }{dt}}
 + {\frac{d\sigma_{^{\uparrow\downarrow }}}{dt}}
 + {\frac{d\sigma_{^{\downarrow\uparrow }} }{dt}}},
\end{equation}
where index $n$ means that spins polarized along a normal to the
scattering plane. Other parameters $A_{ll}$, $A_{ss}$ and $A_{sl}$
have the similar to Eq. (\ref{ann}) structure and differ by the
orientation of spins in the initial state. Rotational invariance
and particle identity leads to the following equality
\begin{equation}\label{srcor}
\Delta\sigma^{el}_T(s)=-4\int_{-s+4m^2}^0
A_{nn}(s,t)\frac{d\sigma}{dt}dt,
\end{equation}
where $\Delta\sigma^{el}_T(s)$ is the cross section difference
with protons polarized along normal to beam direction:
\[
\Delta\sigma^{el}_T(s)\equiv \sigma^{el}_{\uparrow\downarrow}(s)-
\sigma^{el}_{\uparrow\uparrow}(s)
\]
Parity conservation combined with particle identity allows us to
get another relation
\begin{equation}\label{srcoral}
\Delta\sigma^{el}_L(s)=-4\int_{-s+4m^2}^0
A_{ll}(s,t)\frac{d\sigma}{dt}dt,
\end{equation}
where $\Delta\sigma^{el}_L(s)$ is the cross section difference for
the protons polarized along beam direction:
\[
\Delta\sigma^{el}_L(s)\equiv
\sigma^{el}_{{^\rightarrow_\leftarrow}}(s)-
\sigma^{el}_{{^\rightarrow_\rightarrow}}(s)
\]
We also have due to rotational invariance that
\begin{equation}\label{srcoreq}
\int_{-s+4m^2}^0 [A_{nn}(s,t)- A_{ss}(s,t)]\frac{d\sigma}{dt}dt=0,
\end{equation}
And parity
conservation and rotational invariance provide
\begin{equation}\label{srcoral1}
\int_{-s+4m^2}^0 A_{sl}(s,t)\frac{d\sigma}{dt}dt=0.
\end{equation}

Similar relations can be written for the spin correlation
parameters in the inclusive processes.

All above  sum rules should be, of course, in agreement with the
experimental data and therefore they can be used for the
extrapolation to the region where data are absent at the moment.
These sum rules are also interesting as a test ground for the
models  and must be obeyed under their construction.


\small \begin{thebibliography}{9}
\bibitem{bks}
C. Bourrely, E. Leader, J. Soffer, Phys. Rep. 59,  95 1980;\\ S.M.
Troshin, N.E. Tyurin,
 {\it Spin Phenomena in Particle Interactions}, World Scientific Publishing Co., Singapore, 1994;\\
 E. Leader, {\it Spin in Particle Physics}, Cambridge University Press, UK, 2001.
\bibitem{adk}
V.G. Luppov, et al., AIP Conf. Proc. 675,  538, 2003.
\bibitem{kaz}
Yu. M. Kazarinov, et al., Nucl. Phys. B 124,  391, 1977.
\bibitem{bin}
I. Auer, et al., Phys. Lett. 70,  475, 1977;\\
G. Fidecaro, et al., Phys. Lett. 76,  369, 1978;\\
J. Antille, et al., Nucl. Phys. 185,  1, 1981; \\
V.D. Apokin, et al. Sov. Journal of Nucl. Phys. 45,  1355, 1987;\\
D.G. Crabb, et al., Phys. Rev. Lett. 41,  1350, 1978; Phys. Rev.
Lett. 65, 3241, 1990.
\bibitem{osa}
S.M. Troshin, N.E. Tyurin,
     Proceedings ``Polarization Phenomena In
    Nuclear Physics'', Osaka 1985, 954.
\end{thebibliography}
\end{document}